\documentclass[aps,pra,reprint,showkeys,showpacs,amsfonts,amsmath,amssymb,superscriptaddress]{revtex4-1}

\usepackage{graphicx}
\usepackage[latin1]{inputenc}

\begin{document}

\title{Semi-Spectral Method for the Wigner equation}

\author{O. Furtmaier}
\email{oliverfu@ethz.ch}
\affiliation{ETH Z\"{u}rich, Institute for Building Materials, Schafmattstrasse 6 CH-8093 Z\"{u}rich}
\author{S. Succi}
\email{succi@iac.cnr.it}
\affiliation{Istituto per le Applicazioni del Calcolo C.N.R, Via dei Taurini 19, 00185 Rome, Italy}
\author{M. Mendoza}
\email{mmendoza@ethz.ch}
\affiliation{ETH Z\"{u}rich, Institute for Building Materials, Schafmattstrasse 6 CH-8093 Z\"{u}rich}

\begin{abstract}
We propose a numerical method to solve the Wigner equation in quantum systems of spinless, non-relativistic particles. The method uses a spectral decomposition into $L^2(\mathbb{R}^d)$ basis functions in momentum-space to obtain a system of first-order advection-reaction equations. The resulting equations are solved by splitting the reaction and advection steps so as to allow the combination of numerical techniques from quantum mechanics and computational fluid dynamics by identifying the skew-hermitian reaction matrix as a generator of unitary rotations. The method is validated for the case of particles subject to a one-dimensional (an-)harmonic potential using finite-differences for the advection part. 
Thereby, we verify the second order of convergence and observe non-classical 
behavior in the evolution of the Wigner function.   
\end{abstract}


\maketitle

\section{Introduction}
The Wigner formalism, also known as quantum mechanics in phase space \cite{zachos05}, provides an alternative but equivalent description of quantum mechanics in trems of a (quasi)-distribution
function of the particle position and momentum. 
It has proven to provide a helpful supplement to operator methods in Hilbert space 
as well as to path integral formulations, and has offered new insights into the relation between quantum 
and classical physics, as it does not discriminate between coordinate and momentum space. 
For instance, it has been a fruitful perspective for the study of quantum chaos. 
In addition, it offers the opportunity to systematically consider quantum corrections to the classical dynamics by expanding the \textit{quantum Liouville equation} 
around $\hbar \approx 0$. Nowadays, it is also a valuable tool in the fields of quantum optics as well as nuclear, plasma and semiconductor physics to describe transport 
processes, for example, in open quantum systems \cite{zachos05}. 
The Wigner function, introduced by E. Wigner in Ref. \cite{wig32}, is the Weyl transformation 
of the density matrix  and a quasi-probability distribution that can ``intuitively'' account for scattering and decoherence effects in quantum systems \cite{joos96,zurek05}. It differs from a classical 
probability distributions as it can change sign during the evolution especially 
in regions where quantum interference effects become important. 

Since the Wigner equation was introduced in 1932, it has been tackled by various numerical approaches, such as finite differences \cite{frensley90,lee99}, 
Fourier spectral collocation \cite{ring90,arnold95}, deterministic particle \cite{niclot88,wong03}, and Monte-Carlo \cite{dimov14,dimov15}. 
Here, we extend the technique described in Refs. \cite{ring90,arnold95} to arbitrary basis functions $\phi_n(\vec{p})$ of $L^2(\mathbb{R}^d)$ in momentum-space and reveal the 
underlying mathematical structure of the resulting infinite-dimensional set of reaction-advection equations. By using this formulation we show that the action of the potential 
on the Wigner function is a unitary rotation of its coefficient vector, whereas the advection operation can be discretized by various techniques used in computational fluid dynamics, 
such as finite difference, finite volume or finite element, cf. Ref. \cite{loeh08}. In that way, one is able to construct a finite element simulation of the Wigner evolution. 
Employing a more general basis we assume that the higher computational costs of our method, $\mathcal{O}(N^2)$, compared to $\mathcal{O}(N\log N)$ for the spectral Fourier 
decomposition are outweighed by a smaller number $N$ of basis functions to obtain the same order of accuracy through focusing the computational effort to regions of interest, 
as for example in the case of Wigner functions that are strongly localized in momentum-space, such as particles in a periodic potential, cf. Bloch's theorem and Ref. 
\cite{kand98}. In addition, the ``artificial'' periodization of the Wigner function can be avoided, which may mitigate the self-interaction of the distribution at the domain 
boundaries of the simulation that is present for the Fourier basis choice, cf. Ref. \cite{shao11}.

This article is organized as follows. First, we give an introduction to the Wigner formalism and present the properties of the Wigner equation, especially for the 
pseudo-differential operator. In section \ref{numerical_method} we show the details of the numerical method to handle the obtained multi-dimensional reaction-advection equation. 
Thirdly, we validate the technique by simulating a one-dimensional (an-)harmonic oscillator, which offers the opportunity to compare with an analytical solution, cf. Ref. 
\cite{groene46}, such that we can perform a convergence analysis, and to observe quantum effects when the anharmonic potential is used. To study tunneling phenomena, we show the 
evolution of bounded states in the double well potential and measure the spread as well as the covariance of the Wigner function in phase space. 
In the last section, we will highlight the strengths and weaknesses of the approach. 

\section{Wigner formalism}
\label{intro_wig}


Our aim is to simulate the time-evolution of the Wigner function $w(\tau,\vec{q},\vec{p})$ of a $d$-dimensional system of non-relativistic spinless particles of mass $m$ subject
to the potential $U(\tau,\vec{q})$, based on the Wigner equation
\begin{align}
 0 &= \partial_{\tau} w + \frac{\vec{p}}{m}\cdot\vec{\nabla}_q w + \Theta[U]w\text{ ,}\label{wigEQ0}\\
 \Theta[U] &\equiv \frac{\imath}{\hbar}\left[U(\tau,\vec{q}+\imath\hbar\vec{\nabla}_p/2)-U(\tau,\vec{q}-\imath\hbar\vec{\nabla}_p/2)\right]\text{ .}
\end{align}
The independent variables are time $\tau$, space $\vec{q}$ and momentum $\vec{p}$ respectively. Hence, the Wigner function itself has the dimension $h^{-d}$, since it fulfills
\begin{equation}
 \int\limits_{\mathbb{R}^d}\!\mathrm{d}\vec{q} \int\limits_{\mathbb{R}^d}\!\mathrm{d}\vec{p}\hspace{1ex}w(\tau,\vec{q},\vec{p}) = N_p \label{WigNorm}\text{ ,}
\end{equation}
where $N_p$ is the number of particles in the system. To have an easier grasp on the equation, we make it dimensionless. First of all, we measure the Wigner function with respect 
to $\hbar^d$, i.e. we introduce the dimensionless Wigner function $W(\tau,\vec{q},\vec{p})\equiv \hbar^d w(\tau,\vec{q},\vec{p})/N_p$, and we employ the following scaling 
relations
\begin{equation*}
 \vec{x} = \vec{q}/l\text{ , } t = \tau/T\text{ , }\vec{v} = \frac{T}{l m}\vec{p}\text{ , } V(t,\vec{x}) = U(\tau,\vec{q})/\bar{U}\text{ ,}
\end{equation*}
described in Ref. \cite{ring89}. Thus, we obtain the dimensionless Wigner equation 
\begin{align}
 0 &= \partial_t W + \vec{v}\cdot \vec{\nabla}_x W + \Theta[V]W\text{ ,}\label{wigEQ}\\
 \Theta[V] &= \frac{\imath B}{\epsilon}\left[V\left(t,\vec{x}+\frac{\imath\epsilon}{2}\vec{\nabla}_v\right)-V\left(t,\vec{x}-\frac{\imath\epsilon}{2}\vec{\nabla}_v\right)\right]\text{ ,}
\end{align}
where we have introduced the dimensionless constants
\begin{equation}
\epsilon \equiv \frac{\hbar T}{l^2 m}\text{ , } 
B \equiv \frac{\bar{U}T^2}{l^2 m}\text{ ,}
\end{equation}
which we call \textit{effective} Planck's constant and potential strength, respectively. The names are not arbitrary and reflect the natural occurrence of these numbers in the 
dimensionless formulation of the dynamics. For instance, Eq.~\eqref{WigNorm} directly translates into
\begin{equation}
 \epsilon^{-d}\int\limits_{\mathbb{R}^d}\!\mathrm{d}\vec{x} \int\limits_{\mathbb{R}^d}\!\mathrm{d}\vec{v}\hspace{1ex}W(t,\vec{x},\vec{v}) = 1 \text{ ,}
\end{equation}
where we have replaced Planck's constant by its scaled counterpart; the Wigner transform of a pure state $\Psi$ becomes
\begin{equation}
 W = \int\limits_{\mathbb{R}^d}\!\mathrm{d}\vec{y} \hspace{1ex}\Psi^*\left(t,\vec{x}+\frac{\vec{y}}{2}\right)\Psi\left(t,\vec{x}-\frac{\vec{y}}{2}\right)\frac{e^{\imath \vec{v}\cdot\vec{y}/\epsilon}}{(2\pi)^d}\text{ ,}
 \label{Wigner_trafo_2}
\end{equation}
where $\Psi(t,\vec{x})\equiv l^{d/2}\psi(\tau,\vec{q})$ is the dimensionless wave function in position space; the dimensionless time-dependent Schr\"{o}dinger equation reads
\begin{equation}
 \imath\epsilon \partial_t \Psi = \left[|\hat{\vec{v}}|^2/2+B \hat{V}(t,\vec{x})\right]\Psi\text{ ,}
\end{equation}
and the canonical commutation relation is written as
\begin{equation}
 \left[\hat{x}_j,\hat{v}_k\right] = \imath \epsilon \delta_{j,k}\text{ .}
\end{equation}
The symbol $\Theta[V]$ stands for the pseudo-differential operator, whose action on $W$ can be written as an integral
\begin{align}
\Theta[V]W &= \frac{\imath B}{\epsilon}\int\limits_{\mathbb{R}^d}\!\mathrm{d}\vec{\eta}\hspace{1ex}\delta V(t,\vec{x},\vec{\eta}) \hat{W}(t,\vec{x},\vec{\eta})e^{-\imath\vec{\eta}\cdot\vec{v}} \text{ ,} \label{pseudo_analytic}\\
\delta V(t,\vec{x},\vec{\eta}) &\equiv V\left(t,\vec{x}+\frac{\epsilon}{2}\vec{\eta}\right)-V\left(t,\vec{x}-\frac{\epsilon}{2}\vec{\eta}\right)\text{,}\\
\hat{W}(t,\vec{x},\vec{\eta}) &= \frac{1}{(2\pi)^{d}} \int\limits_{\mathbb{R}^d}\!\mathrm{d}\vec{p}\hspace{1ex}W(t,\vec{x},\vec{p})e^{\imath\vec{\eta}\cdot\vec{p}}\text{ .}
\end{align}
If the potential is locally well-approximated by a Taylor series around $\epsilon\approx 0$ we can write 
\begin{equation}
V(t,\vec{x}+\epsilon\vec{\eta}/2) \approx \sum_{|\lambda|=0}^\infty \left(\epsilon/2\right)^{|\lambda|}\frac{D_x^{\lambda}V(t,\vec{x})}{\lambda!} \vec{\eta}^{\lambda}\text{ ,}
    \end{equation}
where $\lambda$ is a multi-index of dimension $d$, i.e. $|\lambda|\equiv \sum_{i=1}^d \lambda_i$, $\lambda ! = \prod_{i=1}^d \lambda_i !$, and $\vec{\eta}^{\lambda} \equiv \prod_{j=1}^d\eta_j^{\lambda_j}$. 
Therefore, the action of the pseudo-differential operator on the Wigner function reads
\begin{equation}
      \Theta[V]W = -B\sum_{|\lambda|\in\mathbb{N}_{odd}}(\imath\epsilon/2)^{|\lambda|-1} \frac{1}{\lambda!}\left(D^{\lambda}_xV\right)\left(D^{\lambda}_v W\right) \text{ .} \label{pseudo_taylor}
\end{equation}
$\mathbb{N}_{odd}$ stands for strictly positive odd integers, such that the sum is always real. For this treatment, the potential needs to be an \textit{analytic function} 
defined on an open set $D\subset \mathbb{R}^d\times\mathbb{R}$, i.e. explicit dependence on time is possible. This implies two important properties for us. Firstly, the function 
is locally given by a convergent power or Taylor series. Secondly, one can find an upper bound for all derivatives of the function, since for every compact set $K\subset D$, for 
all $(t,\vec{x})\in K$, and for all $|\lambda|\in\mathbb{N}_0$ there exists a constant $C$ such that
\begin{equation}
|D^{\lambda}_x V| \leq C^{|\lambda|+1}\lambda! \text{ .}
\end{equation}
Hence, by choosing the right time and length scale ($T,l$) for the problem one can find a convergent power series representation of the pseudo-differential operator. The correct 
choice means 
\begin{equation}
\lim_{|\lambda|\rightarrow \infty} \left(\frac{C\epsilon}{2}\right)^{|\lambda|} = 0 \text{ ,}
\end{equation}
such that the series in Eq. \eqref{pseudo_taylor} converges (locally) uniformly under the assumption that $D_v^{\lambda}W$ is bounded.
\subsection{General properties}
\label{gen_prop}
Expanding the Wigner function in momentum-space into a set of orthonormal basis functions $\{\phi_k\}_{k\in\mathbb{N}}$ of $L^2(\mathbb{R}^d)$ with the inner product
\begin{equation}
 \langle \phi_i, \phi_j \rangle_2 \equiv \int\limits_{\mathbb{R}^d}\!\mathrm{d}\vec{v}\hspace{1ex} \phi^*_i(\vec{v})\phi_j(\vec{v}) = \delta_{i,j} \label{ortho_prop}\text{ ,}
\end{equation}
meaning that
\begin{equation}
 W(t,\vec{x},\vec{v}) = \sum_{k\in\mathbb{N}} a_k(t,\vec{x}) \hspace{1ex}\phi_k(\vec{v}) \text{ ,}
 \label{expan}
\end{equation}
we can rewrite the Wigner equation, Eq. \eqref{wigEQ}, into an infinite system of linear, first-order partial differential equations (PDEs) for the coefficients 
$a_k(t,\vec{x}) \in \mathbb{C}$. The system can be derived by using the orthonormality property of the basis functions, Eq. \eqref{ortho_prop}. Depending on the choice of the 
basis we will find different sets of PDEs. In general, all the sets can be written as a multi-dimensional \textit{reaction-advection} equation
\begin{equation}
 \partial_t \vec{a} + \sum_{i=1}^d A^{(i)} \partial_{x_i} \vec{a} + M_V(t,\vec{x})\vec{a} = \vec{0} \text{ ,}
 \label{num1}
\end{equation}
where $A^{(i)}, M_V(t,\vec{x})$ are square matrices and $\vec{a} = (a_1,a_2,a_3,\dots)$ is the coefficient vector. Independent of the basis choice, the matrix $M_V(t,\vec{x})$ is 
\textit{skew-hermitian}, which can be demonstrated employing Eq. \eqref{pseudo_taylor} or Eq. \eqref{pseudo_analytic}. When using formula \eqref{pseudo_taylor} we have to assume 
that the basis functions are $C^\infty(\mathbb{R}^d)$. Under this condition we can shift the uneven derivatives, $|\lambda|\in\mathbb{N}_{odd}$, which appear as summands in the 
pseudo-differential operator, to show the skew-hermiticity. Demonstrating this property for a \textit{general} set of basis functions of $L^2(\mathbb{R}^d)$, i.e. even 
non-differentiable, we use Eq. \eqref{pseudo_analytic} to write
\begin{align}
 \left(M_V \vec{a}\right)_k =& \int\limits_{\mathbb{R}^d}\!\mathrm{d}\vec{v}\hspace{1ex}\phi^*_k(\vec{v})\left(\Theta[V]W\right)(t,\vec{x},\vec{v})\\
 =& \frac{\imath B}{\epsilon} \int\limits_{\mathbb{R}^d}\!\mathrm{d}\vec{\eta}\hspace{1ex}\delta V(t,\vec{x},\vec{\eta}) \hat{W}(t,\vec{x},\vec{\eta})\notag\\
 &\times\int\limits_{\mathbb{R}^d}\!\mathrm{d}\vec{v}\hspace{1ex}\phi^*_k(\vec{v})e^{-\imath\vec{v}\cdot\vec{\eta}}\text{ ,}
\end{align}
from which we conclude
\begin{align}
 \left(M_V\right)_{k,l} =& \frac{\imath B}{(2\pi)^d\epsilon} \int\limits_{\mathbb{R}^d}\!\mathrm{d}\vec{\eta}\hspace{1ex}\delta V(t,\vec{x},\vec{\eta})\notag\\ 
 &\times\int\limits_{\mathbb{R}^d\times\mathbb{R}^d}\!\mathrm{d}\vec{v}\mathrm{d}\vec{p}\hspace{1ex}\phi^*_k(\vec{v})e^{-\imath\vec{v}\cdot\vec{\eta}}\phi_l(\vec{p})e^{\imath\vec{p}\cdot\vec{\eta}}\label{pseudo_matrix}\text{ .}
\end{align}
This equation confirms the skew-hermiticity of the matrix representation of the pseudo-differential operator, which is an important property for the stability of the proposed 
algorithm as we will see in the next section. An example of this matrix representation is shown in appendix \ref{Pseudo_Matrix}. The entries of the matrix $A^{(i)}$ are given by
\begin{equation}
 \left(A^{(i)}\right)_{k,l} = \int\limits_{\mathbb{R}^d}\!\mathrm{d}\vec{v} \hspace{1ex}\phi^*_k(\vec{v})v_i\phi_l(\vec{v})\text{ ,} \label{StreamMatrix}
\end{equation}
which shows that it is \textit{hermitian}.
\section{Numerical method}
\label{numerical_method}

For the numerical treatment, the expansion in Eq. \eqref{expan}, is cut at the index $N$, i.e. we assume all higher coefficients to be zero. The problem is hence shifted to the 
time-evolution of the $N$-dimensional coefficient vector with the initial condition
\begin{equation}
 \vec{a}(t_0,\vec{x}) = \int\limits_{\mathbb{R}^d}\!\mathrm{d}\vec{v}\hspace{1ex}W(t_0,\vec{x},\vec{v})\vec{\phi}(\vec{v})\text{ ,}
\end{equation}
where $\vec{\phi} = (\phi_1,\phi_2,\dots,\phi_N)$. Therefore, we work with a finite set of $N$ balance equations (PDEs) in the form of Eq. \eqref{num1}. It is important to note 
that, thanks to the Cauchy-Kowaleski theorem, see Ref. \cite{hoerm83}, we know that the system will \textit{locally} have a \textit{unique analytical} solution if the coefficient 
matrix $M_V$ is an analytic function. This condition is sufficient, since the matrices $A^{(i)}$ are constant. In addition, we would like to mention that this does not 
necessarily apply if $M_V$ belongs to the larger group of smooth functions, see Levy's argument in Ref.~\cite{hoerm83}.
\subsection{Operator-splitting}
To proceed with the problem we use an operator-splitting technique (``divide-and-conquer"), i.e. we separate the action of the ``streaming", 
\begin{equation}
\mathcal{S}\vec{a} \equiv -\sum_{i=1}^d A^{(i)} \partial_{x_i} \vec{a} \text{ ,}
\end{equation}
and ``forcing",
\begin{equation}
 \mathcal{F}_t\vec{a}\equiv -M_V(t,\vec{x})\vec{a} \text{ ,}
\end{equation}
operators to apply them sequentially. First, we discretize the time interval from zero to $t$ in $N_t$ periods of duration $\delta t$. Then we can write the approximated solution 
to Eq. \eqref{num1} as
\begin{align}
 \vec{a}(t,\vec{x}) &\approx \overleftarrow{\prod\limits_{k=0}^{N_t-1}} \exp\left(\mathcal{S}\delta t+\int\limits_{k\delta t}^{(k+1)\delta t}\!\mathrm{d}t' \hspace{1ex}\mathcal{F}_{t'}\right)\vec{a}_0(\vec{x})\text{,}\notag\\
  &\approx \overleftarrow{\prod\limits_{k=0}^{N_t-1}} e^{\mathcal{S}\delta t} \exp\left(\int\limits_{k\delta t}^{(k+1)\delta t}\!\mathrm{d}t' \hspace{1ex}\mathcal{F}_{t'}\right)\vec{a}_0(\vec{x})\text{ ,}\notag\\
  &\approx \overleftarrow{\prod\limits_{k=0}^{N_t-1}} e^{\mathcal{S}\delta t} e^{\mathcal{F}_{k\delta t}\delta t}\vec{a}_0(\vec{x})+\mathcal{O}(\delta t) \text{ ,}
\end{align}
where in the first step we have used the third-order accurate Fer expansion \cite{iser99}; in the second step, simple operator splitting; and the numerical integration procedure 
\begin{equation}
 \int\limits_{k\delta t}^{(k+1)\delta t}\!\mathrm{d}t' \hspace{1ex}\mathcal{F}_{t'} \approx \mathcal{F}_{k\delta t}\delta t + \mathcal{O}(\delta t^2)\text{ ,}
\end{equation}
in the third step. It is important to apply the operators in a time-ordered product series, which is indicated by the arrow above the product sign. The obtained method will be 
first-order accurate if it is stable and the numerical procedure for each operator (streaming and forcing) is at least second-order accurate. The total error arises since the 
matrices $A^{(i)}\partial_{x_i}$ and $M_V$ are in general not commuting and because of the second-order accurate integration procedure.
For a second-order accurate method we write
\begin{align}
 \vec{a}^*(t,\vec{x})  &\approx \overleftarrow{\prod\limits_{k=0}^{N_t-1}}e^{\mathcal{S}
\delta t}\exp\left(\int\limits_{k\delta t}^{(k+1)\delta t}\!\mathrm{d}t' \hspace{1ex}\mathcal{F}_{t'}\right)\vec{a}_0^*(\vec{x})\text{,}\label{2ndOrder}
\\ 
 \vec{a}^*(t,\vec{x})&\equiv e^{-\frac{1}{2}\mathcal{F}_{t}\delta t}\vec{a}(t,\vec{x})\text{ ,}\notag
\end{align}
where we have used the Strang-splitting \cite{strang68}. To achieve the demanded accuracy we have to use a third-order accurate integration formula for the forcing operation, 
whereas second-order accuracy in the definition of $a^*$ is sufficient, because it acts only twice during the evolution. If the potential has an explicit time-dependence one can 
use the midpoint rule,  
\begin{equation}
\int\limits_{k\delta t}^{(k+1)\delta t}\!\mathrm{d}t' \hspace{1ex}\mathcal{F}_{t'}\approx \mathcal{F}_{(k+\frac{1}{2})\delta t}\delta t +\mathcal{O}(\delta t^3)\text{ .}
\end{equation}
For the Wigner-Poisson problem \cite{mark90} where one needs to determine the self-consistent electro-static potential, $\Delta V = e\rho(t,\vec{x})$, at every time-step, we make 
use of the fact that the forcing operation does not change the density and hence the electro-static potential. Taking Eq. \eqref{pseudo_taylor} and integrating by parts we can 
show
\begin{equation}
 \int\limits_{\mathbb{R}^d}\!\mathrm{d}\vec{v}\hspace{1ex}\left(\partial_t W + \Theta[V]W\right) = \partial_t \rho = 0 \text{ .}
\end{equation}
Consequently, if the numerical procedure in this step conserves the density up to $\mathcal{O}(\delta t^3)$, it will be sufficient to re-calculate the forcing operator after each 
streaming, which coincides with our time-step definition in Eq. \eqref{2ndOrder}. The questions that remain to be solved are how to compute approximations of the operators' 
actions $e^{\mathcal{S}\delta t}\vec{a}$ (``streaming'') and $e^{\mathcal{F}_k \delta t}\vec{a}$ (``forcing'') such that the resulting algorithm is stable, computationally 
efficient, and of the desired accuracy (first- or second-order). 
\subsection{Forcing}
As it was mentioned in section \ref{gen_prop}, the matrix $M_V$ is skew-hermitian, which means that it belongs to the Lie algebra of the group of unitary matrices. Depending on 
the basis choice we might also find the subgroups of special unitary or special orthogonal matrices if $M_V$ is a traceless, skew-Hermitian, complex matrix or a real, 
skew-symmetric one. Hence, the action of the forcing operator is a \textit{unitary rotation} of the coefficient vector, whose matrix form can in general be calculated before 
starting the simulation. For a skew-symmetric matrix one could use the method described in Ref. \cite{gallier02} or a Pad\'{e} approximation \cite{loan03}
\begin{equation*}
e^{\mathcal{F}_{k\delta t}\delta t} \approx \left[\mathbf{1}+\frac{\delta t}{2}M_V(k\delta t,\vec{x})\right]^{-1}\left[\mathbf{1}-\frac{\delta t}{2}M_V(k\delta t,\vec{x})\right] \text{ .}
\end{equation*} 
For the case of the Wigner-Poisson problem one needs to compute the product of matrix times vector at every time-step, which for instance can be efficiently done with the 
"Expokit" software package \cite{expokit} or using a pre-calculated explicit formula.

\begin{figure}[htbp]
\begin{center}
\input{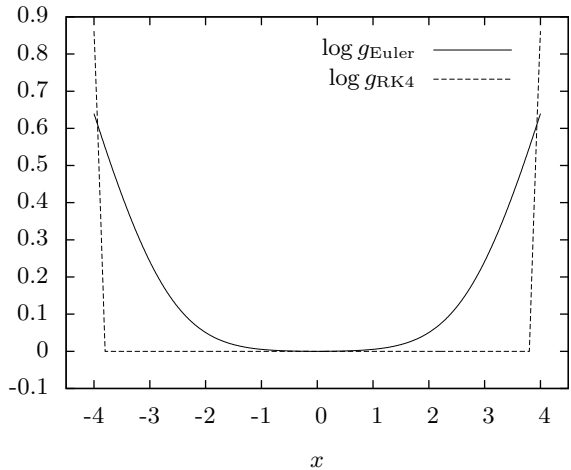}
\caption{Amplification factor for an anharmonic potential ($K=0.5$) using Euler ($1/\delta x = 100$) and RK4 ($1/\delta x = 50$) methods with $N=10$.}
\label{stab_amp}
\end{center}
\end{figure}

\begin{figure}[htbp]
\begin{center}
\input{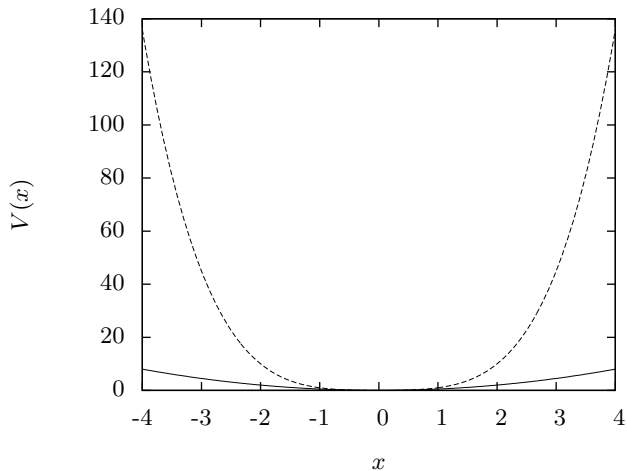}
\caption{Dimensionless anharmonic potential for $K=0$ (solid) and $K=0.5$ (dashed).}
\label{pot_fig}
\end{center}
\end{figure}

One might be tempted to use explicit schemes, such as Euler or Runge-Kutta, to approximate the forcing. However, these methods can become unstable for strongly changing potentials and poor temporal and spatial resolution, which we will show for two examples by evaluating the amplification factor $g$ in von Neumann's stability analysis \cite{keller66}. Consider a time-independent one-dimensional anharmonic potential and the Euler as well as the fourth-order Runge-Kutta method (RK4) as approximations of the forcing, whose amplification factors are given by 
\begin{align}
g_{\text{Euler}} &= |\mathbf{1}-\delta t M_V(x)|_2 \text{ ,}\\
g_{\text{RK4}} &= |\mathbf{1}+\sum_{j=1}^4 \frac{[-\delta t M_V(x)]^j}{j!}|_2\text{ ,}
\end{align}
where $|\dots|_2$ stands for the 2- or spectral-norm, such that Parseval's identity is applicable. 
The plots for both methods are shown in Fig. \ref{stab_amp}. One observes the big amplification factor at the domain boundary, caused by the strong potential variation in this 
area, cf. Fig. \ref{pot_fig}, which may eventually trigger a numerical instability.

\subsection{Streaming}
The streaming can in general be achieved by methods handling (non-)linear hyperbolic systems of conservation laws, often used in computational fluid dynamics, such as finite 
difference, volume, elements or lattice Boltzmann \cite{succi92}. Using, for example, in $d=1$ a flux vector splitting \cite{steger81}, we first diagonalize the matrix $A^{(1)}=TD_AT^{-1}$. Then we define the 
new coefficient vector $\vec{b}(t,\vec{x})\equiv T^{-1}\vec{a}(t,\vec{x})$ and the modified forcing term $\tilde{M}_V \equiv T^{-1}M_V T$, such that the new system of partial 
differential equations can be written as 
\begin{equation}
 \partial_t \vec{b} + D_A \partial_{x} \vec{b} + \tilde{M}_V(t,\vec{x})\vec{b} = \vec{0} \text{ .}
\end{equation}
To simulate the action of the streaming operator, one can now employ the first-order accurate upwind or the second-order accurate Lax-Wendroff scheme \cite{loeh08}, 
since $A^{(1)}$ only has real eigenvalues, i.e. $D_A$ is a real diagonal matrix. The drawbacks of the explicit methods are that the Courant-Friedrichs-Levy condition
\begin{equation}
\frac{|\lambda|_{max}\delta t}{\delta x} \leq 1 \text{ ,}
\end{equation}
needs to be fulfilled for a stable simulation (\textit{conditional stability}) and that they introduce a considerable amount of dissipation, especially if structures with large 
gradients are streamed \cite{loeh08}. If we are employing Hermite functions we could - in the spirit of the lattice Boltzmann method \cite{grad49b,succi01,mendoza10} - use an ``exact" streaming 
operation, which will mitigate the dissipative effects. For this we perform a discrete Hermite transform in $\vec{v}$-space\cite{leibon08} from the coefficient vector to the 
Wigner function, stream, and transform back to the coefficient vector. However, a more detailed analysis and implementation will be the topic 
of a subsequent article.   
  
\subsection{Stability}
The proposed method for evolving the Wigner function will be stable if the operations streaming and forcing are both stable. Since the action of the forcing can be described as a 
unitary rotation one should make sure that the numerical technique conserves this property and hence has an amplification factor of unity. In that respect, the skew-hermiticity 
of $M_V$, the matrix representation of $\Theta[V]$, is a crucial property for the stability of such algorithms, which we will demonstrate in appendix \ref{instability} for an 
asymmetric Hermite basis. This may require a very accurate result for the rotation matrix or the usage of Clifford algebras \cite{snygg97} to perform the rotation. However, the 
least computationally expensive operation is the direct matrix-vector multiplication $\mathcal{O}(N^2)$, whereas the use of the algebra will need slightly more operations 
(although the scaling is the same). For the streaming, one can use any stable method that handles linear advection equations, such as flux vector splitting \cite{steger81}, 
Godunov, finite volume or finite element \cite{loeh08}. The resulting time-evolution of the Wigner function will hence be stable.

\section{Simulation}
For the validation of our numerical procedure we simulate the time-evolution of an (an-)harmonic oscillator. The advantages of these examples are that, on one hand, 
we can compare with the analytical Wigner function of a harmonic oscillator, which is calculated as described in Ref. \cite{groene46}. On the other hand, we can observe 
the effects of quantum corrections to the classical dynamics for an anharmonic potential 
$U_{\text{anh}}(\vec{q})=\frac{1}{2}m\omega |\vec{q}|^2 + \frac{m^2 \omega^3 K}{\hbar}|\vec{q}|^4$ \cite{wig32}. In the case of the double well potential 
$U_{\text{mh}}(\vec{q}) = c m\omega |\vec{q}|^2 + \frac{m^2 \omega^3 K}{\hbar}|\vec{q}|^4$ we can observe the tunneling phenomenon in the Wigner formalism, 
since for certain parameter ranges $c<0$ and $K>0$ the system has states with eigenenergies below $0$ which would not allow classical particles to travel from one potential 
minimum to the other.
As we have described in the introduction to the Wigner formalism, see section \ref{intro_wig}, we use a dimensionless form of the Schr\"{o}dinger and Wigner equations. 
For our examples, we take $l\equiv\sqrt{\frac{\hbar}{m\omega}}$, $T\equiv\frac{1}{\omega}$ and $\bar{U}\equiv\hbar \omega$ as length, time, and potential scales, respectively, 
to find $\epsilon = 1$ and $B=1$. The dimensionless time-dependent Schr\"{o}dinger equation reads
\begin{equation}
\imath\partial_t \Psi = \left(\frac{|\hat{\vec{v}}|^2}{2}+c|\hat{\vec{x}}|^2+K|\hat{\vec{x}}|^4\right)\Psi\text{ ,}
\end{equation}
such that the eigenfunctions and -values of the dimensionless Hamilton operator at $K=0$ and $c=1/2$ are given by
\begin{align}
\Psi^{|n|}_n(\vec{x}) &= \frac{e^{-|\vec{x}|^2/2}}{\sqrt{\pi^{d/2} 2^{|n|} n!}}H^{|n|}_n(\vec{x})\text{ ,}\\
\mathcal{E}_n &= |n|+d/2\text{ ,}
\end{align}
where $n=(n_1,\dots,n_d)$ is a multi-index and
\begin{equation}
H^{|n|}_n(\vec{x}) = (-1)^{|n|}e^{|\vec{x}|^2}\left(D^k e^{-|\vec{x}|^2}\right)\text{ ,}
\end{equation}
the $d$-dimensional Hermite polynomial, according to Ref. \cite{grad49a}. The dimensionless Wigner equation in differential form becomes
\begin{equation}
\partial_t W + \vec{v}\cdot\vec{\nabla}_x W -2(c+2K|\vec{x}|^2) \vec{x}\cdot\vec{\nabla}_v W + \Theta_c[K]W = 0\text{ ,}
\label{WigHarm}
\end{equation}
where 
\begin{equation}
\Theta_c[K] W\equiv \frac{K}{4} \sum_{|\lambda|=3}\frac{D_x^{\lambda}|\vec{x}|^4}{\lambda !}D^{\lambda}_v W
\end{equation}
is the quantum correction to the ``classical'' dynamics of the particle.
\subsection{Basis of Hermite functions}
In our simulation, we choose Hermite functions as orthonormal basis set in momentum-space, i.e.
\begin{equation}
 \phi_k^{|k|}(\vec{v}) = \frac{e^{-|\vec{v}|^2/2}}{\sqrt{\pi^{d/2} 2^{|k|} k!}}H^{|k|}_k(\vec{v})\text{ ,}\label{multi_Hermite}
\end{equation}
where $k$ is a multi-index of dimension $d$. Hence, Eq. \eqref{expan} changes to
\begin{equation}
 W(t,\vec{x},\vec{v})=\sum_{|k|=0}^N a^{|k|}_k(t,\vec{x}) \hspace{1ex}\phi^{|k|}_k(\vec{v})\text{ .}
\end{equation}
The number of basis functions that is needed to simulate the evolution of a given state will in general depend on how wide the spread of the corresponding Wigner
function is in momentum space. However, by scaling the Hermite functions, cf. Ref. \cite{hollo98}, one can significantly reduce $N$ to simulate eigenstates with higher energy.
In order to find the necessary number of basis functions for a chosen accuracy one needs to take a look at the variation of the resulting Wigner function with respect to changes 
in $N$.
In order to find the initial coefficients, $\vec{a}(t_0,\vec{x})$, we use the property of the Hermite polynomials or Hermite functions, defined by Eq. \eqref{multi_Hermite}, 
that they diagonalize the Fourier transform operator, 
\begin{equation}
\int\limits_{\mathbb{R}^d}\!\mathrm{d}\vec{v}\hspace{1ex}e^{\imath\vec{y}\cdot\vec{v}/\epsilon}\phi^{|k|}_k(\vec{v}) = (\sqrt{2\pi})^d \imath^{|k|}\phi_k^{|k|}(\vec{y})\text{ .}
\end{equation}
The proof is given in Ref. \cite{groene46}. Using the Wigner transform, defined by Eq. \eqref{Wigner_trafo_2}, we can write
\begin{align*}
 a^{|k|}_k(t_0,\vec{x}) &= \int\limits_{\mathbb{R}^d}\!\mathrm{d}\vec{v}\hspace{1ex}W(t_0,\vec{x},\vec{v})\phi^{|k|}_k(\vec{v})\text{ ,}\\
 &= \int\limits_{\mathbb{R}^d}\!\mathrm{d}\vec{y}\hspace{1ex}\Psi^*\left(t_0,\vec{x}+\frac{\epsilon\vec{y}}{2}\right)\Psi\left(t_0,\vec{x}-\frac{\epsilon\vec{y}}{2}\right)\phi_k^{|k|}(\vec{y})\\
 &\times \frac{\epsilon^d \imath^{|k|}}{(2\pi)^{\frac{d}{2}}}\text{ .}
\end{align*}
One can see that the obtained coefficients are real due to the symmetry properties of the Hermite functions $\phi_k^{|k|}(-\vec{y}) = (-1)^{|k|}\phi_k^{|k|}(\vec{y})$. In addition, we can simplify Eq. \eqref{pseudo_matrix} by using the same property to obtain
\begin{equation}
 \left(M_V\right)_{k,l} = \frac{B}{\epsilon}\imath^{|k|-|l|-1}\int_{\mathbb{R}^d}\!\mathrm{d}\vec{\eta}\hspace{1ex}\delta V(t,\vec{x},\vec{\eta})\phi_k^{|k|}(\vec{\eta})\phi_l^{|l|}(\vec{\eta})\text{ .}
 \label{pseudo_matrix_harm}
\end{equation}
Looking at this result, one can realize that $M_V$ is a real, skew-symmetric matrix. The matrices $A^{(i)}$ are real, symmetric and \textit{sparse} for this basis choice. 
They are sparse, because, regardless how the basis functions are ordered at most two entries per row or column are non-zero due to the recursion relation of the $d$-dimensional 
Hermite polynomials \cite{grad49a}. For further explanations on the conservation and convergence properties for this basis choice, we refer the reader to Ref. \cite{hollo98}, 
where the authors treat the Vlasov equation, which can be considered as the classical limit, $\epsilon\rightarrow 0$, of the Wigner equation.

\subsection{Harmonic oscillator}
\begin{figure}[htbp]
\begin{center}
\input{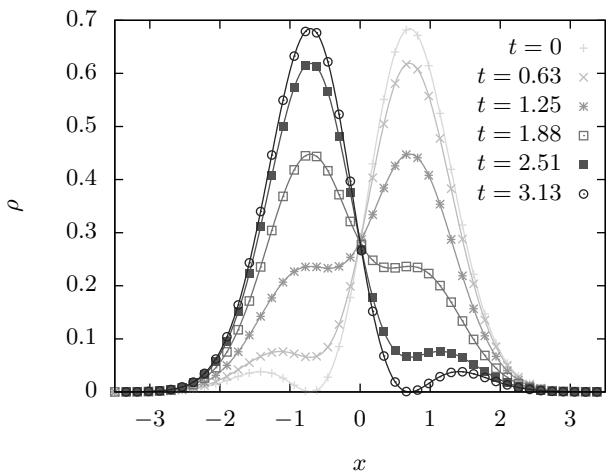}
\caption{Temporal evolution of the probability density for the harmonic potential ($c=0.5$, $K=0$), $\rho_{\Psi}$ (solid lines) and $\rho_W$ (points)
for $\Psi=(\Psi_0+\Psi_1)/\sqrt{2}$ using $N=16$ Hermite basis functions.}
\label{densHarm_evol}
\end{center}
\end{figure}

\begin{figure}[htbp]
\begin{center}
\input{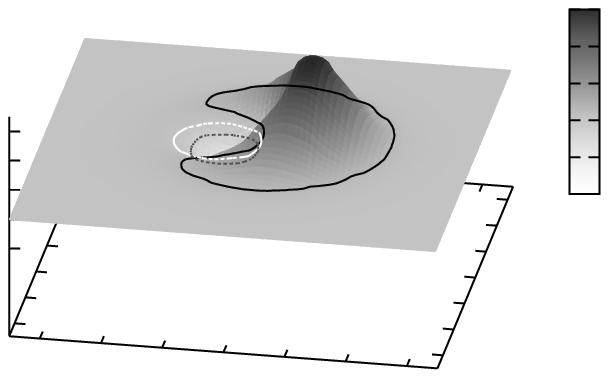}
\caption{Wigner function for the harmonic potential ($c=0.5$, $K=0$) at $t=0$ for the superposition $\Psi=(\Psi_0+\Psi_1)/\sqrt{2}$ using $N=16$ Hermite basis functions; 
contour lines at $W=0$ (white) and $W=\pm0.025$ (black/gray).}
\label{wigHarm0}
\end{center}
\end{figure}

\begin{figure}[htbp]
\begin{center}
\input{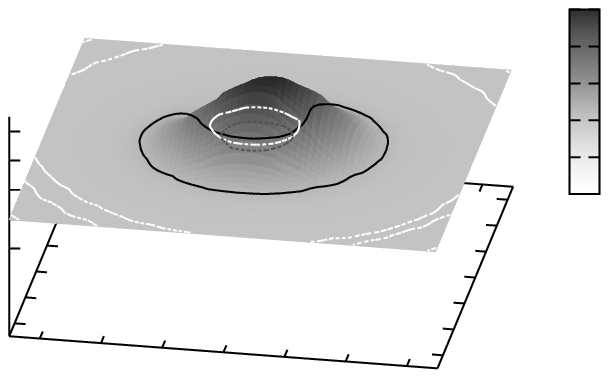}
\caption{Wigner function for the harmonic potential ($c=0.5$, $K=0$) at $t=1.25$ for the superposition $\Psi=(\Psi_0+\Psi_1)/\sqrt{2}$ using $N=16$ Hermite basis functions; 
contour lines at $W=0$ (white) and $W=\pm0.025$ (black/gray).}
\label{wigHarm2}
\end{center}
\end{figure}

\begin{figure}[htbp]
\begin{center}
\input{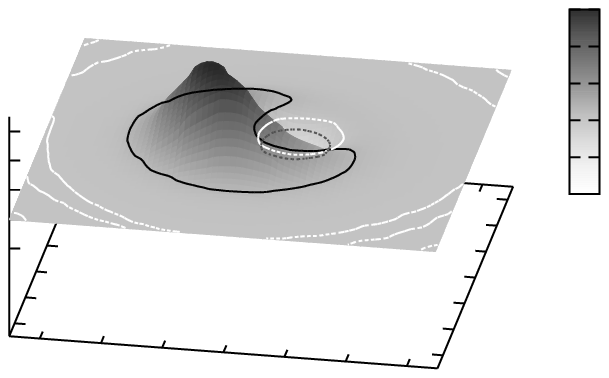}
\caption{Wigner function for the harmonic potential ($c=0.5$, $K=0$) at $t=2.51$ for the superposition $\Psi=(\Psi_0+\Psi_1)/\sqrt{2}$ using $N=16$ Hermite basis functions; 
contour lines at $W=0$ (white) and $W=\pm 0.025$ (black/gray).}
\label{wigHarm4}
\end{center}
\end{figure}

\begin{figure}[htbp]
\begin{center}
\input{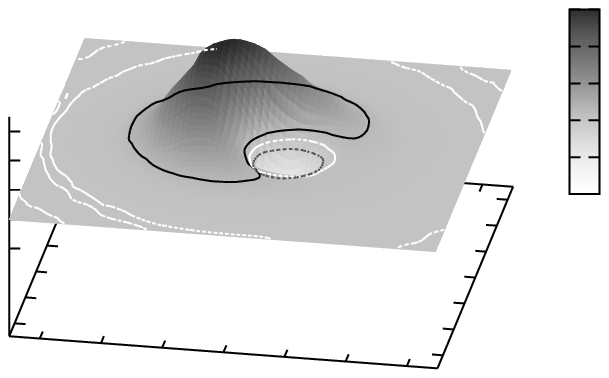}
\caption{Wigner function for the harmonic potential ($c=0.5$, $K=0$) at $t=3.76$ for the superposition $\Psi=(\Psi_0+\Psi_1)/\sqrt{2}$ using $N=16$ Hermite basis functions; 
contour lines at $W=0$ (white) and $W=\pm0.025$ (black/gray).}
\label{wigHarm6}
\end{center}
\end{figure}

\begin{figure}[htbp]
\begin{center}
\input{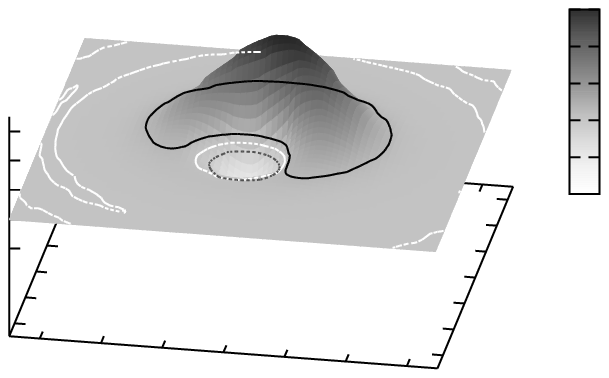}
\caption{Wigner function for the harmonic potential ($c=0.5$, $K=0$) at $t=5.01$ for the superposition $\Psi=(\Psi_0+\Psi_1)/\sqrt{2}$ using $N=16$ Hermite basis functions; 
contour lines at $W=0$ (white) and $W=\pm0.025$ (black/gray).}
\label{wigHarm8}
\end{center}
\end{figure}

We run a simulation of an one-dimensional harmonic oscillator with the second order accurate method, using a Lax-Wendroff scheme for the streaming, a spatial resolution of $\delta x = 1/50$, and periodic boundaries at $x=\pm 3.5$. The resulting matrices for the reaction-advection system can be calculated using formula \eqref{pseudo_taylor_matrix} in appendix \ref{Pseudo_Matrix}. As initial state $\Psi$ we choose a superposition between ground and first excited state $\frac{\Psi_0+\Psi_1}{\sqrt{2}}$, since the Wigner function of a single eigenstate is time-independent. Thus, we can observe the evolution for a system whose probability density changes in time. In Fig. \ref{densHarm_evol}, we show a comparison between the analytical spatial probability density $\rho_{\Psi}(t,x)\equiv|\Psi(t,x)|^2$ and the probability density calculated from the Wigner function with
\begin{equation*}
 \rho_W(t,x) \equiv \int\limits_{\mathbb{R}}\!\mathrm{d}v\hspace{1ex}W(t,x,v) = \sum_{k=0}^N a_k(t,x)\int\limits_{\mathbb{R}}\!\mathrm{d}v\hspace{1ex}\phi_k(v)\text{ .}
\end{equation*}
The comparison shows very good agreement. However, the actual results for the Wigner function are more insightful, 
since they contain additional information. They are shown in Figs. \ref{wigHarm0}-\ref{wigHarm8} together with the contour lines at $W=0$ and $W=\pm0.025$. 
One observes a ``rigid'' rotation of the Wigner function in phase space, which is typical for the harmonic oscillator. This can be seen by using the method of characteristics for 
solving Eq. \eqref{WigHarm} at $K=0$ and $c=0.5$ which leads to solving the Hamilton equations
\begin{equation}
 \dot{x} = v \text{ , } \dot{v} = -x\text{ .}
\end{equation}
The period for one revolution is $T=2\pi(\mathcal{E}_1-\mathcal{E}_0)=2\pi$, which is also confirmed by the simulation in terms of the temporal error margin. The contour line at $W=0$ close to the boundaries shows patterns which are not present in the analytical solution. They are caused by the numerical error fluctuations, see Fig. \ref{WigHarmDev}, since the magnitude of the Wigner function in that region becomes comparable to the numerical error.

\subsection{Convergence}
\begin{figure}[htbp]
\begin{center}
\input{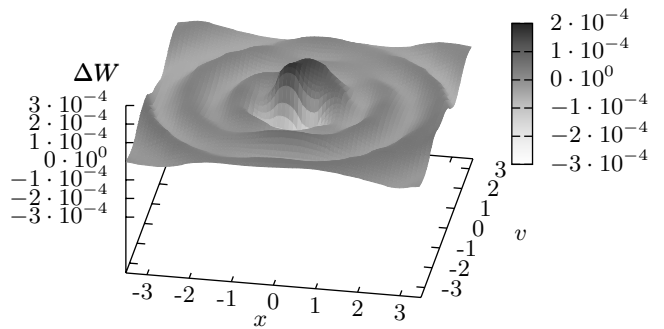}
\caption{Error of the Wigner function for the harmonic potential ($c=0.5$, $K=0.5$) at $t=6.28$ for the superposition $\Psi=(\Psi_0+\Psi_1)/\sqrt{2}$ with $N=16$, $\delta x = 1/50$.}
\label{WigHarmDev}
\end{center}
\end{figure}

\begin{figure}[htbp]
\begin{center}
\input{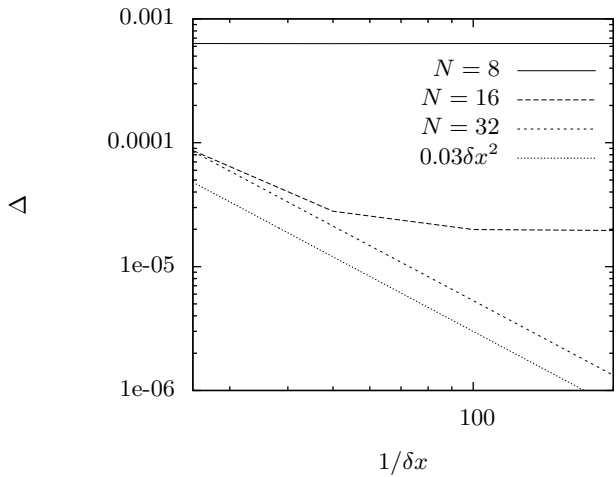}
\caption{Convergence analysis of the harmonic Wigner function after $t=2\pi$ with respect to $\delta x$ and $N_b$.}
\label{2ndconv}
\end{center}
\end{figure}

Based on the work in Ref. \cite{groene46}, one can calculate the exact Wigner transform $W_{\text{ex}}$ of any wave function $\Psi(t,\vec{x})$ expanded in Hermite functions. 
We will use this formula to calculate the Wigner transform for an eigenstate $\Psi_n(t,x)$ of the harmonic oscillator and compare our results for different numbers of basis functions $N$ and spatial resolutions $1/\delta x$. It is important to note that the exact Wigner function of an eigenstate for $K=0$, $c=0.5$ and $d=1$ is given by Laguerre polynomials through
\begin{align}
W_n(x,v) &= \frac{(-1)^{n}}{\pi}L_n[2(x^2+v^2)]e^{-x^2-v^2}\text{ ,}\\
L_n(y) &\equiv \frac{1}{n!}\left(\frac{\mathrm{d}}{\mathrm{d}x}-1\right)^n x^n\text{ ,}
\end{align} 
which does not give a finite expansion into Hermite functions in $v$. The deviation of our results from the analytical solution after one period is shown in Fig. \ref{WigHarmDev}. We observe that the error is of the order of $10^{-4}$ and its magnitude is rather homogeneously distributed. In Fig. \ref{2ndconv} we show the convergence of the second order accurate method by looking at the error 
\begin{align}
 \Delta &\equiv \sqrt{\frac{1}{N_x N_v}\sum_{i,j} |\Delta W(x_i,v_j,t)|^2}\text{ ,}\\
 \Delta W(x_i,v_j,t) &\equiv W(x_i,v_j,t)-W_{\text{ex}}(x_i,v_j,t)\text{ ,}
\end{align}
for periodic boundary conditions in real-space and a domain size $x\in[-5,5]$. The error is evaluated by choosing the same momentum- and space-grid. The domain size is chosen such that boundary effects do not significantly influence the error in the convergence analysis, since $W(t,\pm 5,v)\sim \mathcal{O}(10^{-10})$. Looking at Fig. \ref{2ndconv}, we observe that the second order convergence can only be verified for sufficiently many basis functions (here: $N= 32$). This behavior is caused by a total error that is composed by the discretization of time and real-space as well as the approximation of the Wigner function with a finite number of basis functions in momentum-space. Therefore, we expect $\Delta$ to saturate for a fixed resolution and an increasing number of basis functions, or in the opposite scenario, which can be deduced from Fig. \ref{2ndconv} for $N=16$. 

\subsection{Anharmonic oscillator}
\begin{figure}[htbp]
\begin{center}
\input{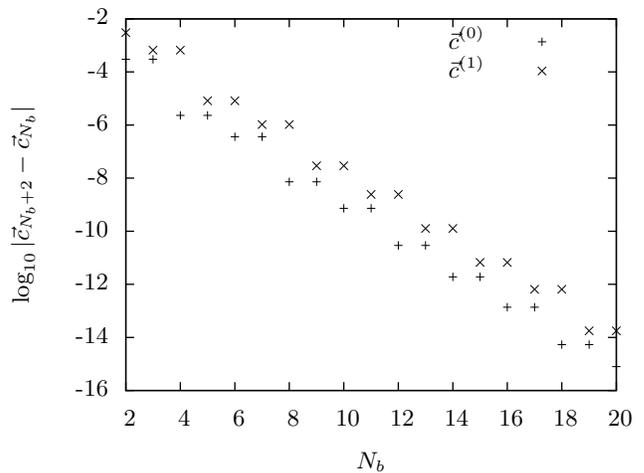}
\caption{Convergence of eigenstate coefficient vector for ground and first excited states of the anharmonic potential ($c=0.5$, $K=0.001$) up to \textit{double precision}.}
\label{conv_K_3}
\end{center}
\end{figure}

\begin{figure}[htbp]
\begin{center}
\input{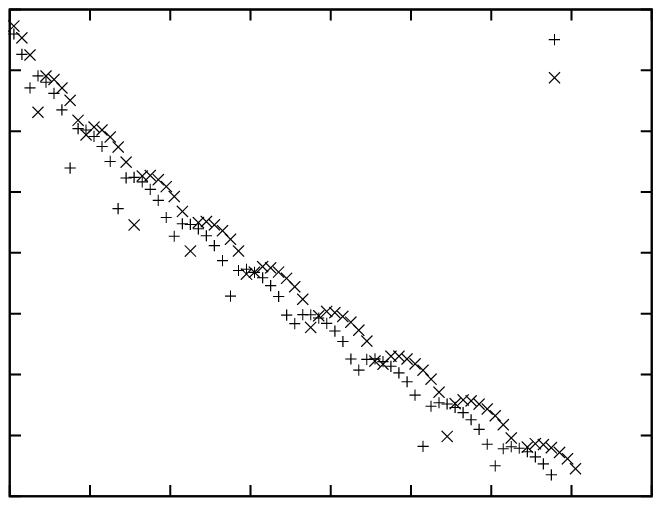}
\caption{Convergence of eigenstate coefficient vector for ground and first excited states of the anharmonic potential ($c=0.5$, $K=0.5$) up to \textit{double precision}.}
\label{conv_K_1}
\end{center}
\end{figure}

For an anharmonic potential $c=0.5$ and $K>0$ we approximate the eigenstates $\Psi_n^{(an)}$ by a superposition of $N_b$ harmonic eigenstates, i.e. 
\begin{equation}
 \Psi_n^{(an)}(t,x) \approx e^{-\imath t \mathcal{E}_n^{(an)}} \sum_{k=0}^{N_b} c^{(n)}_k \Psi_k(x)\text{ .}
 \label{eigen_approx}
\end{equation}
Then we determine the coefficient vector $\vec{c}^{(n)}$ by diagonalizing the matrix representation of the anharmonic Hamilton operator. 
This works very well for moderate anharmonicities, but becomes very costly for $K>10^{-3}$, as can be seen in Figs. \ref{conv_K_3} and \ref{conv_K_1}. 
In addition, one also observes that, as expected, the ground state converges faster than the first excited state.

\begin{figure}[htbp]
\begin{center}
\input{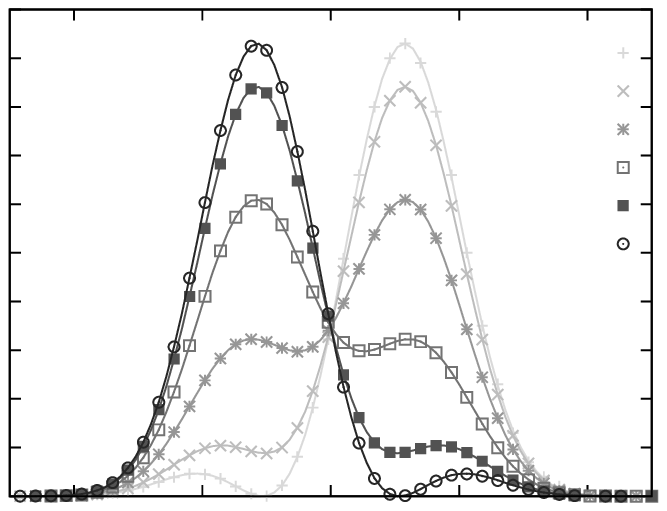}
\caption{Temporal evolution of the probability density for the anharmonic potential ($c=0.5$, $K=0.5$), $\rho_{\Psi}$ (solid lines) and $\rho_W$ (points)
for $\Psi=(\Psi^{(an)}_0+\Psi^{(an)}_1)/\sqrt{2}$ using $N_b=150$, and $N=16$ Hermite basis functions.}
\label{densAnharm_evol}
\end{center}
\end{figure}

\begin{figure}[htbp]
\begin{center}
\input{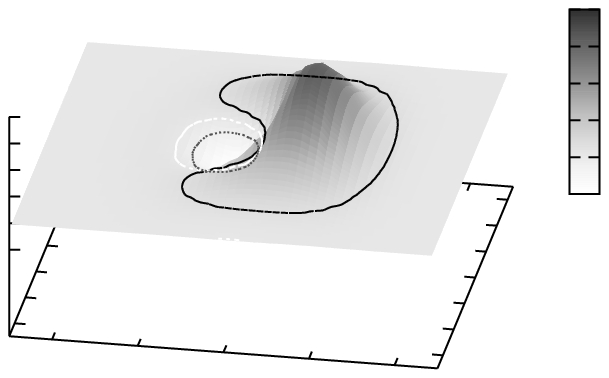}
\caption{Wigner function for the anharmonic potential ($c=0.5$, $K=0.5$) at $t=0$ for the superposition $\Psi=(\Psi^{(an)}_0+\Psi^{(an)}_1)/\sqrt{2}$ using $N_b=150$, and $N=16$ Hermite basis functions; 
contour lines at $W=0$ (white) and $W=\pm0.025$ (black/gray).}
\label{wigAnharm0}
\end{center}
\end{figure}

\begin{figure}[htbp]
\begin{center}
\input{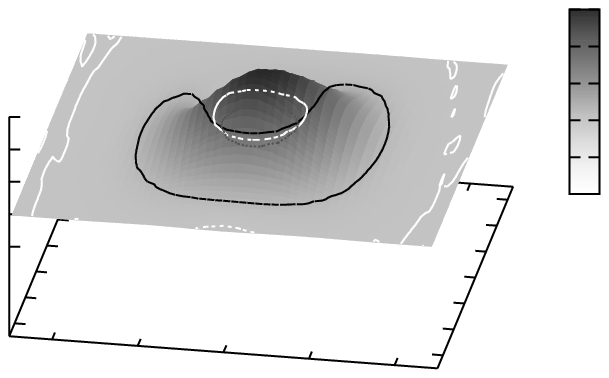}
\caption{Wigner function for the anharmonic potential ($c=0.5$, $K=0.5$) at $t=0.77$ for the superposition $\Psi=(\Psi^{(an)}_0+\Psi^{(an)}_1)/\sqrt{2}$ using $N_b=150$, and $N=16$ Hermite basis functions; 
contour lines at $W=0$ (white) and $W=\pm0.025$ (black/gray).}
\label{wigAnharm2}
\end{center}
\end{figure}

\begin{figure}[htbp]
\begin{center}
\input{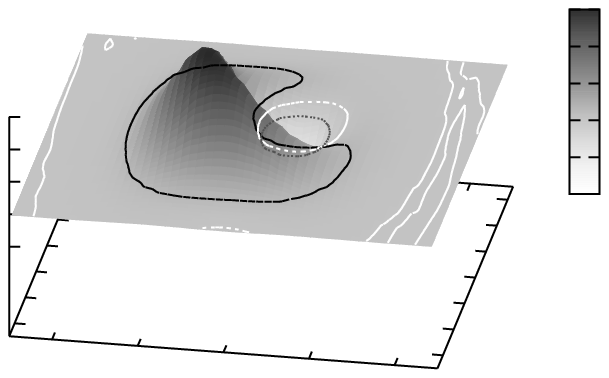}
\caption{Wigner function for the anharmonic potential ($c=0.5$, $K=0.5$) at $t=1.54$ for the superposition $\Psi=(\Psi^{(an)}_0+\Psi^{(an)}_1)/\sqrt{2}$ using $N_b=150$, and $N=16$ Hermite basis functions; 
contour lines at $W=0$ (white) and $W=\pm0.025$ (black/gray).}
\label{wigAnharm4}
\end{center}
\end{figure}

\begin{figure}[htbp]
\begin{center}
\input{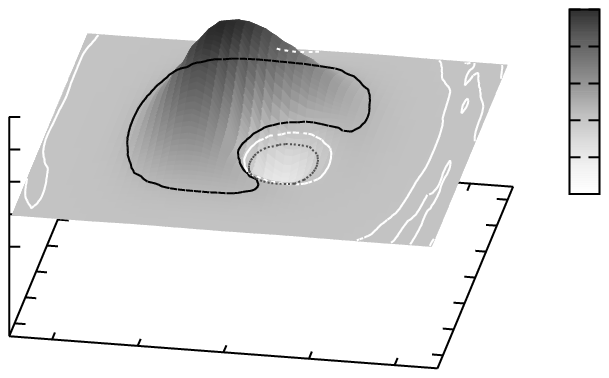}
\caption{Wigner function for the anharmonic potential ($c=0.5$, $K=0.5$) at $t=2.23$ for the superposition $\Psi=(\Psi^{(an)}_0+\Psi^{(an)}_1)/\sqrt{2}$ using $N_b=150$, and $N=16$ Hermite basis functions; 
contour lines at $W=0$ (white) and $W=\pm0.025$ (black/gray).}
\label{wigAnharm6}
\end{center}
\end{figure}

\begin{figure}[htbp]
\begin{center}
\input{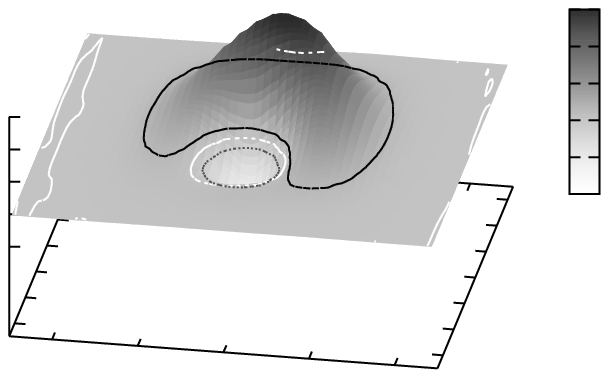}
\caption{Wigner function for the anharmonic potential ($c=0.5$, $K=0.5$) at $t=3.09$ for the superposition $\Psi=(\Psi^{(an)}_0+\Psi^{(an)}_1)/\sqrt{2}$ using $N_b=150$, and $N=16$ Hermite basis functions; 
contour lines at $W=0$ (white) and $W=\pm0.025$ (black/gray).}
\label{wigAnharm8}
\end{center}
\end{figure}

The simulation is run with the second order accurate method for periodic boundary conditions at $x=\pm 3.5$ with a spatial resolution of $1/\delta x = 50$. 
The result for the spatial probability evolution is shown in Fig. \ref{densAnharm_evol}, where we observe a good agreement with the wave function dynamics. 
In Figs. \ref{wigAnharm0}-\ref{wigAnharm8} we show the Wigner function evolution together with the contour lines at $W=0$ and $W=\pm0.025$. 
They depict a ``rotation'' with a smaller period $T_{an}=2\pi/(\mathcal{E}_1^{(an)}-\mathcal{E}_0^{(an)})<2\pi$. In this case it is not a rigid rotation, 
since the Wigner function gets compressed in position- and broadened in momentum-space due to the larger potential and the particle number conservation.  
The contour line at $W=0$ close to the boundaries shows again the numerical error fluctuations, since in this region the magnitude of the Wigner function becomes comparable to the error, 
which is $\mathcal{O}(10^{-4})$. It was estimated by comparing the initial and final Wigner function of 
one revolution.

\subsection{Double well potential}
\begin{figure}[htbp]
\begin{center}
\input{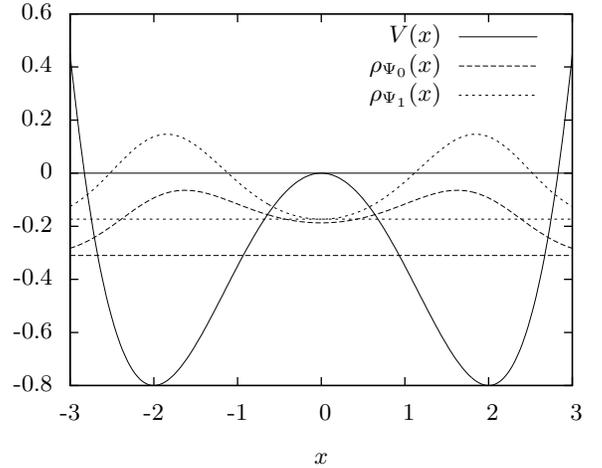}
\caption{Double well potential ($c=-0.4$, $K=0.05$) and probability density of ground and first excited states displaced from $0$ by 
$\mathcal{E}^{(mh)}_0 = -0.310$ (dashed horizontal line) and $\mathcal{E}^{(mh)}_1 = -0.173$ (dotted horizontal line) for $N_b = 86$ Hermite basis functions.}
\label{mexhat}
\end{center}
\end{figure}

\begin{figure}[htbp]
\begin{center}
\input{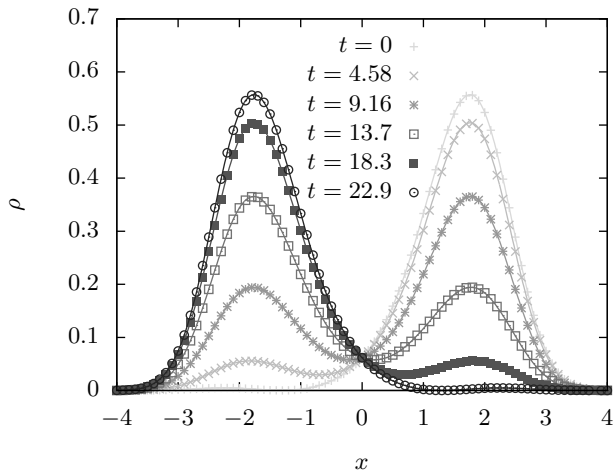}
\caption{Temporal evolution of the probability density for the double well potential ($c=-0.4$, $K=0.05$), $\rho_{\Psi}$ (solid lines) and $\rho_W$ (points)
for the superposition $\Psi=(\Psi^{(mh)}_0+\Psi^{(mh)}_1)/\sqrt{2}$ using $N_b=86$, and $N=32$ Hermite basis functions.}
\label{densMexHat_evol}
\end{center}
\end{figure}

\begin{figure}[htbp]
\begin{center}
\input{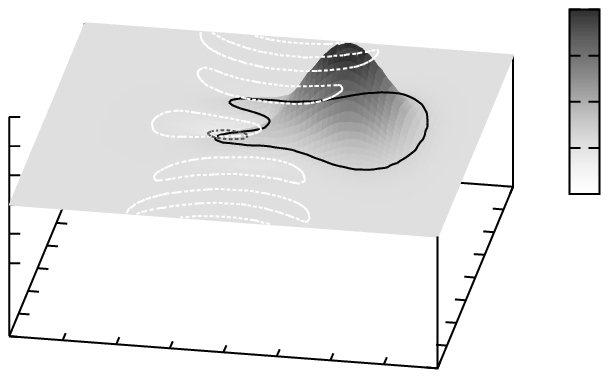}
\caption{Wigner function for the double well potential ($c=-0.4$, $K=0.05$) at $t=0$ for the superposition $\Psi=(\Psi^{(mh)}_0+\Psi^{(mh)}_1)/\sqrt{2}$ using $N_b=86$, and $N=32$ Hermite basis functions; 
contour lines at $W=0$ (white) and $W=\pm0.025$ (black/gray).}
\label{wigMex0}
\end{center}
\end{figure}

\begin{figure}[htbp]
\begin{center}
\input{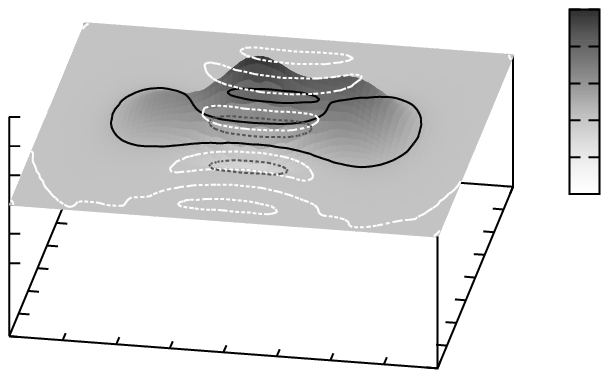}
\caption{Wigner function for the double well potential ($c=-0.4$, $K=0.05$) at $t=9.16$ for the superposition $\Psi=(\Psi^{(mh)}_0+\Psi^{(mh)}_1)/\sqrt{2}$ using $N_b=86$, and $N=32$ Hermite basis functions; 
contour lines at $W=0$ (white) and $W=\pm0.025$ (black/gray).}
\label{wigMex2}
\end{center}
\end{figure}

\begin{figure}[htbp]
\begin{center}
\input{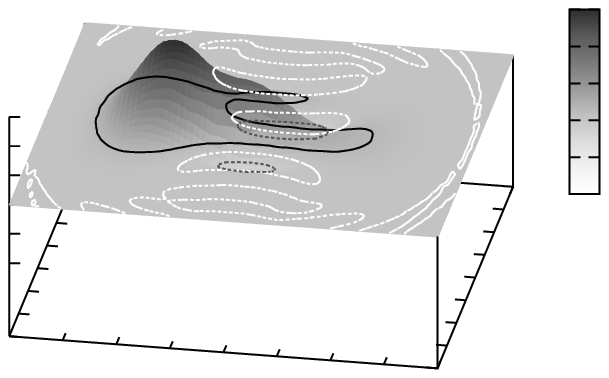}
\caption{Wigner function for the double well potential ($c=-0.4$, $K=0.05$) at $t=18.3$ for the superposition $\Psi=(\Psi^{(mh)}_0+\Psi^{(mh)}_1)/\sqrt{2}$ using $N_b=86$, and $N=32$ Hermite basis functions; 
contour lines at $W=0$ (white) and $W=\pm0.025$ (black/gray).}
\label{wigMex4}
\end{center}
\end{figure}

\begin{figure}[htbp]
\begin{center}
\input{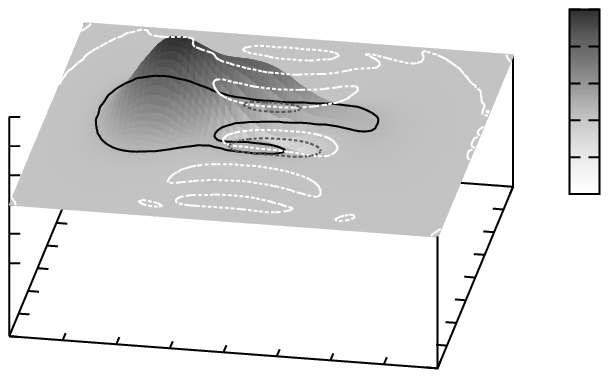}
\caption{Wigner function for the double well potential ($c=-0.4$, $K=0.05$) at $t=27.5$ for the superposition $\Psi=(\Psi^{(mh)}_0+\Psi^{(mh)}_1)/\sqrt{2}$ using $N_b=86$, and $N=32$ Hermite basis functions; 
contour lines at $W=0$ (white) and $W=\pm0.025$ (black/gray).}
\label{wigMex6}
\end{center}
\end{figure}

\begin{figure}[htbp]
\begin{center}
\input{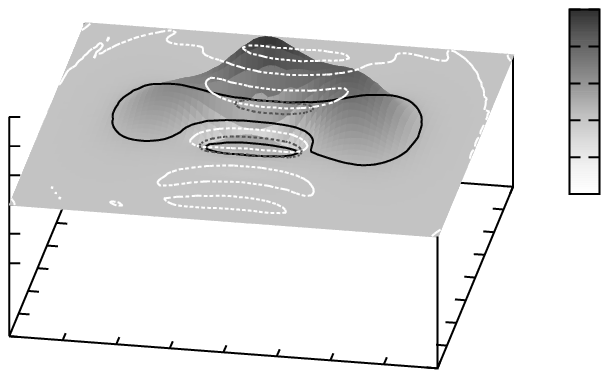}
\caption{Wigner function for the double well potential ($c=-0.4$, $K=0.05$) at $t=36.7$ for the superposition $\Psi=(\Psi^{(mh)}_0+\Psi^{(mh)}_1)/\sqrt{2}$ using $N_b=86$, and $N=32$ Hermite basis functions; 
contour lines at $W=0$ (white) and $W=\pm0.025$ (black/gray).}
\label{wigMex8}
\end{center}
\end{figure}

\begin{figure}[htbp]
\begin{center}
\input{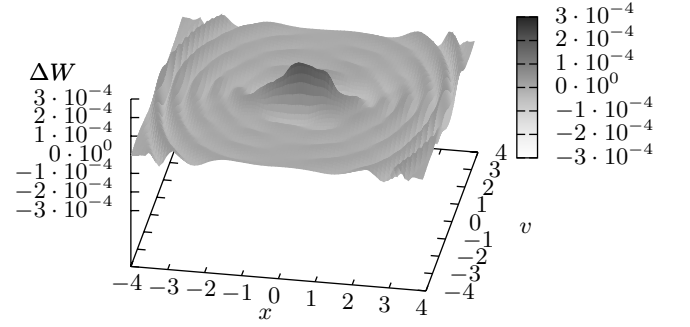}
\caption{Error of the simulated Wigner function for the double well potential ($c=-0.4$, $K=0.05$) at $t=45.9$ for the superposition $\Psi=(\Psi^{(mh)}_0+\Psi^{(mh)}_1)/\sqrt{2}$ 
using $N_b=86$, and $N=32$ Hermite basis functions.}
\label{wigErrorMexHat}
\end{center}
\end{figure}

\begin{figure}[htbp]
\begin{center}
\input{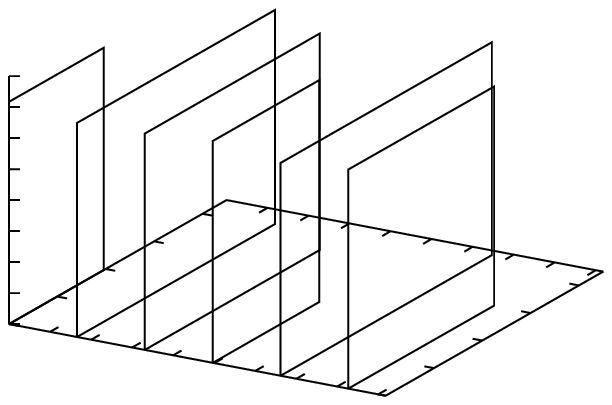}
\caption{Temporal evolution of the spread in the form of rectangles (width $\equiv\Delta x$, height $\equiv\Delta v$, area $\equiv \Delta x\Delta v$) of the simulated Wigner function for the double well potential ($c=-0.4$, $K=0.05$) for the superposition 
$\Psi=(\Psi^{(mh)}_0+\Psi^{(mh)}_1)/\sqrt{2}$ using $N_b=86$, and $N=32$ Hermite basis functions.}
\label{sprevol}
\end{center}
\end{figure}

\begin{figure}[htbp]
\begin{center}
\input{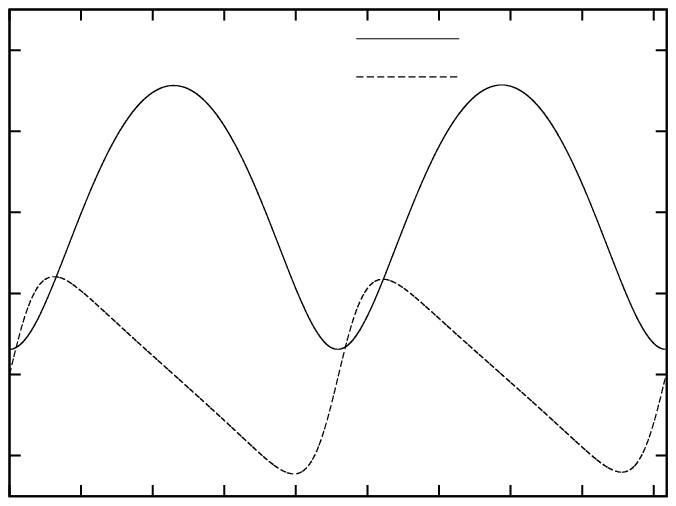}
\caption{Temporal evolution of the uncertainty (solid) and covariance (dashed) of the simulated Wigner function for the double well potential ($c=-0.4$, $K=0.05$) for the superposition 
$\Psi=(\Psi^{(mh)}_0+\Psi^{(mh)}_1)/\sqrt{2}$ using $N_b=86$, and $N=32$ Hermite basis functions.}
\label{covevol}
\end{center}
\end{figure}

To study tunneling effects, we simulate the time-evolution of the Wigner function for ``bounded'' states of a one-dimensional double well potential $V(x)=c x^2 + K x^4$, 
cf. Fig. \ref{mexhat}, with the second order accurate method, periodic boundary conditions at $x=\pm 4$ and spatial resolution $1/\delta x = 50$. We call a state bounded 
if its eigenenergy is smaller than zero and hence below the potential barrier around $x=0$. Taking again the Hermite basis to approximate the ground and first excited state as in 
Eq. \eqref{eigen_approx}, we show in Fig. \ref{densMexHat_evol} the evolution of the probability density in comparison to the evolution according to Schr\"{o}dinger's equation. 
The agreement is very good. In Figs. \ref{wigMex0}-\ref{wigMex8} one can see the evolution of the Wigner function for the tunneling of the state through the potential barrier 
together with contour lines at $W=0$ and $W=\pm0.025$. The error during the revolution is at most $\mathcal{O}(10^{-4})$, as can be seen in Fig. \ref{wigErrorMexHat}. 
In addition, one observes by looking at the contour line for $W=-0.025$ the appearance of ripples and valleys in the front and the back of the positive quasi-probability density during the tunneling process 
of the particle through the potential barrier, which indicate the non-classical behavior in the corresponding coordinate space.
The contour line at $W=0$ close to the boundaries shows again the numerical error fluctuations in regions where the magnitude of the Wigner function becomes comparable to the numerical error. 
The period of the revolution $T_{mh} = 2\pi/(\mathcal{E}^{(mh)}_1-\mathcal{E}_0^{(mh)})\gg 2\pi$ is much larger than the one for a harmonic oscillator.
In addition to the probability density and the Wigner function, we have also analyzed the spread of the Wigner function in phase space by measuring the expectation values 
\begin{align}
(\Delta x^2)(\Delta v^2) &\equiv \langle (\hat{x}-\langle \hat{x}\rangle)^2\rangle\langle (\hat{v}-\langle v\rangle)^2\rangle  \\
&= \left(\langle x^2\rangle_W-\langle x\rangle_W^2\right)\left(\langle v^2\rangle_W-\langle v\rangle_W^2\right)\text{ ,}\\
 \text{Cov}(x,v) &\equiv \left(\frac{1}{2}\langle \hat{x}\hat{v}+\hat{v}\hat{x}\rangle - \langle \hat{x}\rangle\langle \hat{v}\rangle\right)/(\Delta x \Delta v)\\
 &=\frac{\langle xv \rangle_W - \langle x \rangle_W \langle v \rangle_W}{\sqrt{\langle x^2\rangle_W-\langle x\rangle_W^2}\sqrt{\langle v^2\rangle_W-\langle v\rangle_W^2}}\text{ ,}
\end{align}
where $\langle f(x,v)\rangle_W \equiv \int\!\mathrm{d}x\mathrm{d}v \hspace{1ex}f(x,v)W$. The first quantity measures the well-known standard deviation of a 
quantum state in coordinate and momentum space that fulfills Heisenberg's uncertainty principle $\Delta x \Delta v \geq \frac{\epsilon}{2}$. In Fig. \ref{sprevol} we show
the coordinate and momentum uncertainty in the form of rectangles, i.e. the width, height and aread correspond to $\Delta x$, $\Delta v$ and $\Delta x \Delta v$, respectively. 
In that way, one can see that the standard deviation in position measurements mainly contributes to the uncertainty and its temporal change. The second quantity Cov$(x,v)$
is the covariance between the coordinate and momentum variable in the corresponding Wigner function normalized with the standard deviations, such that $|\text{Cov}(x,v)|\leq 1$.
The evolution of these expectaton values is shown in Fig. \ref{covevol}. One observes a periodic behavior with $T=T_{mh}/2$ and finds the maximum uncertainty $\Delta x\Delta v$ 
exactly when the peak of the spatial probability density 
tunnels through the potential barrier in the middle of the double well potential. In contrast Cov$(x,v)$ behaves similar to the first temporal derivative of the uncertainty.

\section{Conclusions}
We have developed a semi-spectral simulation method for the time-evolution of the Wigner quasi-probability distribution that uses a spectral-decomposition of the distribution into 
arbitrary basis functions of $L^2(\mathbb{R}^d)$ in momentum-space, which transforms the original partial differential equation into an infinite-dimensional set of advection-reaction  equations. 

For the numerical treatment, we introduce a cutoff in the expansion, which makes the system finite-dimensional, and split the operators for the reaction and advection part so as 
to apply them sequentially to the distribution function. 
We demonstrated that, due to the skew-hermitian symmetry of the matrix representation 
of the pseudo-differential operator (Lie algebra), the action of the forcing or reaction operator (Lie group) is a unitary rotation, which stabilizes the simulation even for strongly varying potentials compared to other 
explicit methods, such as Euler or RK4. 
The advection or streaming part can be handled by many numerical approaches 
from computational fluid dynamics. Here, we have chosen a flux-vector splitting for the validation of our method by simulating a single, non-relativistic, spinless particle subject to 
a one-dimensional (an-)harmonic or double well potential with Hermite basis functions. 
Having the exact Wigner function of the harmonic oscillator, we verified the second-order 
convergence of the method and also demonstrated its applicability to non-classical dynamics 
in the case of strong anharmonicities and tunneling phenomena. 

The disadvantage of an arbitary basis choice is the higher computational cost of $\mathcal{O}(N^2)$ compared to $\mathcal{O}(N\log N)$ for a Fourier basis, since the pseudo-differential
operator is diagonal for this basis choice, as shown in Refs. \cite{arnold95,ring90} . 
However, if one only considers momentum-derivatives up to order $N_{\lambda} \ll N$ and an 
explicit scheme such as fourth order Runge-Kutta is used, the computational cost also scales like $\mathcal{O}(N)$ \cite{hollo98}. In addition, the artificial periodization of the 
Wigner distribution in momentum-space, caused by the plane wave approximation, lives in a different function space than the original Wigner function, thus giving rise to unphysical 
self-interactions at the domain boundaries \cite{shao11}. 
These basis functions are also not well suited to the simulation of structures that are 
strongly localized in momentum-space,  such as particles in periodic potentials, since 
this would require a very large number of such functions.

The CPU time for the time evolution of one revolution for the harmonic Wigner function,
using the second-order accurate method with $N=32$ Hermite basis functions and a spatial resolution $1/\delta x = 100$, i.e. $700$ grid points and $4558$ time-steps, 
is approximately $7.92\text{ s}$ using a single core of a $3$ GHz Intel(R) Core(TM)2 Quad CPU Q9650 processor.

As future work, we plan to study phase transitions in \textit{open quantum systems}, the effects of scattering (``\textit{quantum Boltzmann equation}''), for example in the case of 
electrons and phonons in semiconductor devices, the effects of boundary conditions, cf. Ref. \cite{frensley90}, and stochastic perturbations. Furthermore, we plan to analyse the influence 
of decoherence on the topology of the Wigner function in two dimensions, cf. Ref. \cite{steuer13}.
%

\begin{acknowledgments}
Financial support from the European Research Council (ERC) Advanced Grant 319968-FlowCCS is kindly acknowledged.
\end{acknowledgments}

\appendix

\section{Matrix-representation of pseudo-differential operator}
\label{Pseudo_Matrix}
For a one-dimensional, analytical potential and the Hermite function basis the matrix-representation of the pseudo-differential operator simplifies from Eq. \eqref{pseudo_matrix_harm}
to 
\begin{align}
 M_V &= \sum_{n=0}^\infty \left(\frac{\epsilon}{2}\right)^{2n} M_n\partial_x^{2n+1}V(t,x) \text{ ,}\label{pseudo_taylor_matrix}\\
 (M_n)_{k,l} &\equiv \imath^{k-l-1}\int\limits_{\mathbb{R}}\!\mathrm{d}\eta\hspace{1ex}\frac{\eta^{2n+1}}{(2n+1)!}\phi_k(\eta)\phi_l(\eta)\text{ .}\notag
\end{align}
Looking at this result one observes how the contributions from odd higher order derivatives scale with the effective Planck constant and the change in sign. The examples for $N=5$ shows the filling of higher order matrices with more and more entries. In the case of $N=5$, the matrix $M_V$ is already filled for $n=2$, i.e. considering the fifth derivative of the potential. As soon as $M_V$ is completely filled an explicit method, such as Euler or Runge-Kutta will have a computational complexity of $\mathcal{O}(N^2)$, although the prefactor will be smaller than for the method which uses the corresponding rotation matrix.
\begin{align}
 M_0 &= \left(
\begin{array}{cccccc}
 0 & -\frac{1}{\sqrt{2}} & 0 & 0 & 0 & 0 \\
 \frac{1}{\sqrt{2}} & 0 & -1 & 0 & 0 & 0 \\
 0 & 1 & 0 & -\sqrt{\frac{3}{2}} & 0 & 0 \\
 0 & 0 & \sqrt{\frac{3}{2}} & 0 & -\sqrt{2} & 0 \\
 0 & 0 & 0 & \sqrt{2} & 0 & -\sqrt{\frac{5}{2}} \\
 0 & 0 & 0 & 0 & \sqrt{\frac{5}{2}} & 0 \\
\end{array}
\right)\notag\\
M_1 &= \left(
\begin{array}{cccccc}
 0 & -\frac{1}{4 \sqrt{2}} & 0 & \frac{1}{4 \sqrt{3}} & 0 & 0 \\
 \frac{1}{4 \sqrt{2}} & 0 & -\frac{1}{2} & 0 & \frac{1}{2 \sqrt{3}} & 0 \\
 0 & \frac{1}{2} & 0 & -\frac{3 \sqrt{\frac{3}{2}}}{4} & 0 & \frac{\sqrt{\frac{5}{6}}}{2} \\
 -\frac{1}{4 \sqrt{3}} & 0 & \frac{3 \sqrt{\frac{3}{2}}}{4} & 0 & -\sqrt{2} & 0 \\
 0 & -\frac{1}{2 \sqrt{3}} & 0 & \sqrt{2} & 0 & -\frac{5 \sqrt{\frac{5}{2}}}{4} \\
 0 & 0 & -\frac{\sqrt{\frac{5}{6}}}{2} & 0 & \frac{5 \sqrt{\frac{5}{2}}}{4} & 0 \\
\end{array}
\right)\notag\\
M_2 &= \left(
\begin{array}{cccccc}
 0 & -\frac{1}{32 \sqrt{2}} & 0 & \frac{1}{16 \sqrt{3}} & 0 & -\frac{1}{16 \sqrt{15}} \\
 \frac{1}{32 \sqrt{2}} & 0 & -\frac{3}{32} & 0 & \frac{\sqrt{3}}{16} & 0 \\
 0 & \frac{3}{32} & 0 & -\frac{19}{32 \sqrt{6}} & 0 & \frac{\sqrt{\frac{5}{6}}}{4} \\
 -\frac{1}{16 \sqrt{3}} & 0 & \frac{19}{32 \sqrt{6}} & 0 & -\frac{11}{16 \sqrt{2}} & 0 \\
 0 & -\frac{\sqrt{3}}{16} & 0 & \frac{11}{16 \sqrt{2}} & 0 & -\frac{17 \sqrt{\frac{5}{2}}}{32} \\
 \frac{1}{16 \sqrt{15}} & 0 & -\frac{\sqrt{\frac{5}{6}}}{4} & 0 & \frac{17 \sqrt{\frac{5}{2}}}{32} & 0 \\
\end{array}
\right)\notag
\end{align}



\section{Example for unstable ``basis'' choice}
\label{instability}
Assuming we are dealing with a harmonic potential, then the quantum corrections vanish and the Wigner and Vlasov equation are identical.
Using an asymmetric Hermite basis, as described in \cite{hollo96}, i.e. 
\begin{equation}
 W(t,x,v) = \frac{e^{-v^2}}{\pi^{1/4}}\sum_{k=0}^N a_k(t,x)\frac{H_k(v)}{\sqrt{2^n n!}} \text{ ,}
\end{equation}
the matrix representation of $\Theta[V]$ will be lower triangular, which can be seen using formula \eqref{pseudo_taylor} and integration by parts. For $N=4$ we find
\begin{equation*}
 M_V(x) = -x\left(
 \begin{array}{ccccc}
           0 & 0 & 0 & 0 & 0 \\
           \sqrt{2} & 0 & 0 & 0 & 0 \\
           0 & 2 & 0 & 0 & 0 \\
           0 & 0 & \sqrt{6} & 0 & 0 \\
           0 & 0 & 0 & 2\sqrt{2} & 0 \\
 \end{array}\right)
\end{equation*}
which is not skew-hermitian or -symmetric anymore. Examining the resulting exact forcing action, we find
\begin{equation*}
 e^{-M_V(x)\delta t} =
\begin{pmatrix}
 1 & 0 & 0 & 0 & 0 \\
 \sqrt{2} x \text{$\delta $t} & 1 & 0 & 0 & 0 \\
 \sqrt{2} x^2 \text{$\delta $t}^2 & 2 x \text{$\delta $t} & 1 & 0 & 0 \\
 \frac{2 x^3 \text{$\delta $t}^3}{\sqrt{3}} & \sqrt{6} x^2 \text{$\delta $t}^2 & \sqrt{6} x \text{$\delta $t} & 1 & 0 \\
 \sqrt{\frac{2}{3}} x^4 \text{$\delta $t}^4 & \frac{4 x^3 \text{$\delta $t}^3}{\sqrt{3}} & 2 \sqrt{3} x^2 \text{$\delta $t}^2 & 2 \sqrt{2} x \text{$\delta $t} & 1 \\
\end{pmatrix}
\end{equation*}
from which we conclude for the amplification factor
\begin{equation*}
 g_{\text{AS}}=\left|e^{-M_V(x)\delta t}\right|_2 > 1 \text{ if } \delta t>0 \text{ , } x\neq 0 \text{ .}
\end{equation*}
This means that the method will become unstable at a certain time in the evolution, as described in Ref. \cite{hollo98}. There are ways to tackle this problem by introducing a 
collision operator, see Ref. \cite{joyce71}, or a filtering technique, further explained in Ref. \cite{cheng76}.

\end{document}